\documentclass[submission, Phys]{SciPost}

\hypersetup{
    colorlinks,
    linkcolor={red!50!black},
    citecolor={blue!50!black},
    urlcolor={blue!80!black}
}

\usepackage[bitstream-charter]{mathdesign}
\usepackage{pstricks,epsfig}
\DeclareSymbolFont{usualmathcal}{OMS}{cmsy}{m}{n}
\DeclareSymbolFontAlphabet{\mathcal}{usualmathcal}

\begin{document}

\begin{center}{\Large \textbf{ Monte Carlos for tau lepton --  Standard Model \\ and New Physics signatures.
\\
}}\end{center}

\begin{center}
Zbigniew Was
\end{center}

\begin{center}
{\bf } IFJ PAN \\ Radzikowskiego 152 \\31-342 Cracow, Poland

zbigniew.was@ifj.edu.pl
\end{center}

\begin{center}
\today
\end{center}


\definecolor{palegray}{gray}{0.95}
\begin{center}
\colorbox{palegray}{
  \begin{tabular}{rr}
  \begin{minipage}{0.1\textwidth}
    \includegraphics[width=30mm]{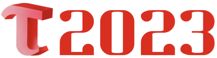}
  \end{minipage}
  &
  \begin{minipage}{0.81\textwidth}
    \begin{center}
    {\it The 17th International Workshop on Tau Lepton Physics}\\
    {\it Louisville, USA, 4-8 December 2023} \\
    \doi{10.21468/SciPostPhysProc.?}\\
    \end{center}
  \end{minipage}
\end{tabular}
}
\end{center}

\section*{Abstract}
{\bf
  One of the purposes of High Energy accelerator experiments
         is  confrontation of theory and measurements in ever new realms.
Any new agreement extends theory applicability domain, any discrepancy  hints to unexplained, calling for  better calculations or new, deeper theory. 
         Often
         one has to search for  small contributions  over  large Standard Model background.
         Multidimensional signatures and complex background subtraction cuts imply that Monte Carlo techniques
         are indispensable.
 My talk included
  description of: KKMC Monte Carlo for $e^+e^- \to \tau^+\tau^- (n\gamma)$ (with $\tau$ decays),
  generation of additional lepton pairs of SM and New Physics, and contributions
   from anomalous magnetic and electric dipole moments.

}


\section{Introduction}
\label{sec:intro}
My talk is devoted to  {\tt KKMC} Precision Monte Carlo  for $e^+e^- \to \tau^+\tau^- (n\gamma)$ , with $\tau$ decays included, and its associated projects.
The main  design purposes, as in the past, include:  (i) precision simulation of lepton pair production, (ii) flexible set up
for initialization of user defined hadronic currents used in $\tau$ decays.
Recent developments in  {\tt KKMC } are particularly important for high energy applications: the {\tt C++}
version of the software is now available and includes improvements over the
{\tt FORTRAN} version, such as better arrangements for beam-spread.
 But the most important aspect was to preserve continuity of
    numerical tests,  an essential ingredient of work see Ref.~\cite{Jadach:2022mbe}. The long road toward    
   enhanced precision of QED part started already ~\cite{Jadach:2023pka}, 
    but this  road toward third order matrix elements is long and perilous.
    There is a need for 
    new contributors to enter, to preserve   skills and  continuity of the projects,
    and at the same time to address new challenges.

    The {\tt C++} version of {\tt KKMC} is public, but is not yet adapted to lower energies  (the
    function for photon vacuum polarization $\Pi_{\gamma\gamma}(s)$ to be obtained 
    from dispersion relations is clearly defined, but needs to be then adjusted).
    There are clear manpower  issues: help in installations, set-ups and of some phenomenology applications is
    not as robust as it was in the past.
    For the {\tt F77} version as installed in Belle II I take care and help, but main burden is now on somebodys else's
    shoulders.

 Pair emission is a necessary step toward third order matrix element installation.
    From the algorithmic point of view, treated as radiative corrections it is incompatible with exponentiation, because
    the corresponding crude distribution does not feature conformal symmetry. This symmetry is essential in building a relation between crude level Monte Carlo with matrix elements of eikonal form. It is also important for matching
    line-shape distribution (intermediate resonances)  and  initial state bremsstrahlung. Alternative solutions are necessary.
    Real pair emission can be achieved by running of {\tt KKMC}  with  appropriate flags for pair
    emission part of
    loop vertex correction switched on. Then  a simultaneous run, of four fermion final state Monte Carlo program, like
    {\tt KORALW} \cite{Jadach:2001mp}, is necessary.
    Alternatively, for final state pair emissions, Monte Carlo of the after-burn type, {\tt PHOTOS} is available
    \cite{Davidson:2010ew}. The program enables full phase-space coverage. Although tests with Matrix element simulations
    indicate that precision is sufficient for today's applications,
  some studies of related  systematic ambiguities are included in Ref.~\cite{Kusina:2022zfb}.
  These publications do not match the FCC precision, and  further work is needed.
  
  In parallel, some extensions with the beyond Standard Model physics are available.
     For $\tau$ decays see \cite{Antropov:2019ald},
    and for production process, contribution to the previous  TAU2021 conference \cite{Banerjee:2021rtn}.
    The  emission of dark photons (dark scalars) decaying to light lepton pairs is addressed.
  Extensions of {\tt KKMC}  Monte Carlo for four fermion final states (processes mediated by $ZH$ boson pair) is
    a possible start for future, long,  development path  to  FCC applicability.  

Another pursued development direction is extension of  simulation for  anomalous magnetic and electric dipole moments.
Recent  g-2 measurements, discussed in other talks of the conference, initiate   interests in anomalous
dipole moments of $\tau$ lepton pair production. 
The question arises how to simulate their impact on differential distributions. We addressed this point with
the help of
algorithms enabling calculation of event weights: ratios of matrix elements for production and decay of
tau lepton pairs
    with other  one, without dipole moment effects.
    First it was done for lower energies $e^+e^- \to \tau^+\tau^-$ processes, as of Belle II,   \cite{Banerjee:2022sgf},
    later for higher energies, as of $Z$ peak and above \cite{Banerjee:2023qjc}.
    In this reference  elements of calculations necessary  for  $ p p  \to \tau^+\tau^- X$ processes are discussed as well.

    The process of migration from {\tt FORTRAN} to {\tt C++} is advancing well. The {\tt C++}  Versions of {\tt PHOTOS}
    Monte Carlo for radiative corrections in decays \cite{Davidson:2010ew}
      and {\tt KKMC}   \cite{Jadach:2022mbe}   
      are already published. The
      {\tt TAUOLA} Monte Carlo  for $\tau$ decays is prepared for such transformation too, but actual implementation
      is not completed.  Its internal structure was already prepared
      some time ago \cite{Chrzaszcz:2016fte}. Further improvements are now  in Belle II hands. Fits and evaluation of ambiguities of explored data are necessary.
    The program was prepared to run with user provided hadronic currents, which could be coded in {\tt C++} and
      activated with redefinition of pointers.
      That solution was temporarily but  partly abandoned. Step back was due to requirements of Belle software organization.
      
      My experience with {\tt PHOTOS}, where migration to {\tt C++},  has been completed  is that the best is to translate program with young researchers' participation.
      They can observe that migration does not go
      too fast, because existing arrangements for future and for tests could  be then easily lost.
      Participation of researchers with distinct profile is helpful,
      and prepares us for future takeover of the projects, independent of whoever it may be.

      The software for Belle II is working well, and I am involved personally. Important contributor  is now
      Swagato Banerjee.
     Installation and initialization for other platforms have recently posed  problems
     especially for {\tt C++}  version of KKMC. For example proper loading of {\tt root} library was
     bringing difficulties, and it was reported to us, but our response was slow.
    That is one of the most visible aspect of the manpower issue for our community,
    but passing expertise to the next generation is potentially larger issue, even if related challenges are not
    imminent to see.
    
    I will skip or cover most of these  points briefly only, and  will focus my presentation mostly
    on anomalous dipole moments. Let me present some details in the following sections.

    \section{Basics of precision Monte Carlo}
\begin{figure}[h]
     \centering
    { \resizebox*{0.49\textwidth}{0.39\textheight}{\includegraphics{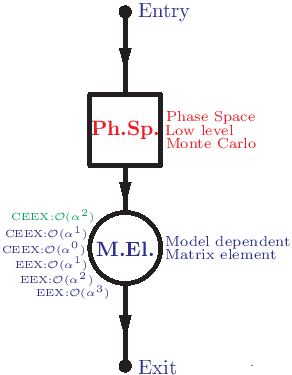}} } 
\caption{Structure of Matrix Element based Monte Carlo programs, in case of {\tt KKMC} Plot is taken  from Ref.~\cite{Jadach:2000ir}.}
\label{Fig:ref}
\end{figure}

    Let me stress at first 
    {\tt KKMC} is based on ``matrix element $\times$ full and exact phase space''.
    Neither {\tt KKMC} nor {\tt PHOTOS} employ a shower-like algorithm. They are
    both non-Markovian. Generation starts with Poisson distribution
    in number of photon candidates. In case of {\tt KKMC} tower of
    simplified/improvable structures starts from  one-dimensional spectrum, then
    an internal generator exploiting eikonal 
    matrix elements  valid all over full multi-photon phase space is devised.
    It was a highly non-trivial step to achieve.
    Next step is exploiting bremsstrahlung part of QED matrix elements and Yennie–Frautschi–Suura  exponentiation \cite{Yennie:1961ad}. That is not all that is needed:
    the  electroweak and strong interaction parts need to be included as well.
    Finally, light lepton  
    pair emissions must be taken into account. I have omitted the complexity of
    presamplers necessary due to collinear configurations, which are naturally
    regulated by electron (or other lepton) masses.
    Finally,  the library of Matrix Elements  devised as
    input for event weight,  sometimes called ``model weight'',
    represents an independent module see Fig. \ref{Fig:ref}. The term ``model weight'' may be 
    misunderstood. There is no room for arbitrary modeling. Amplitudes in use
    are the result of formal fixed order matrix elements. For details see
    \cite{Jadach:2000ir}.
    
    Note that for phase space organization,
    in {\tt KKMC} design, conformal symmetry of multi-photon phase space and
    eikonal parts of QED amplitudes is fundamental. That is why, algorithm
    is suitable for many applications, path for implementation of fixed order
    amplitudes is defined. At present, second order matrix elements are
    installed. In case of {\tt PHOTOS} different solution is used for exact
    multi-photon phase space generation. It is restricted to final state
    radiations only, and solution is not explored beyond use of the first
    order matrix elements. On the other hand, emission of massive particles
    was easier to implement. For details seem \cite{Kusina:2022zfb,Davidson:2010ew} and references therein.

\section{Additional pairs}
\label{sec:another}
\begin{figure}[h]
  \begin{center}
\includegraphics[width=0.44\textwidth]{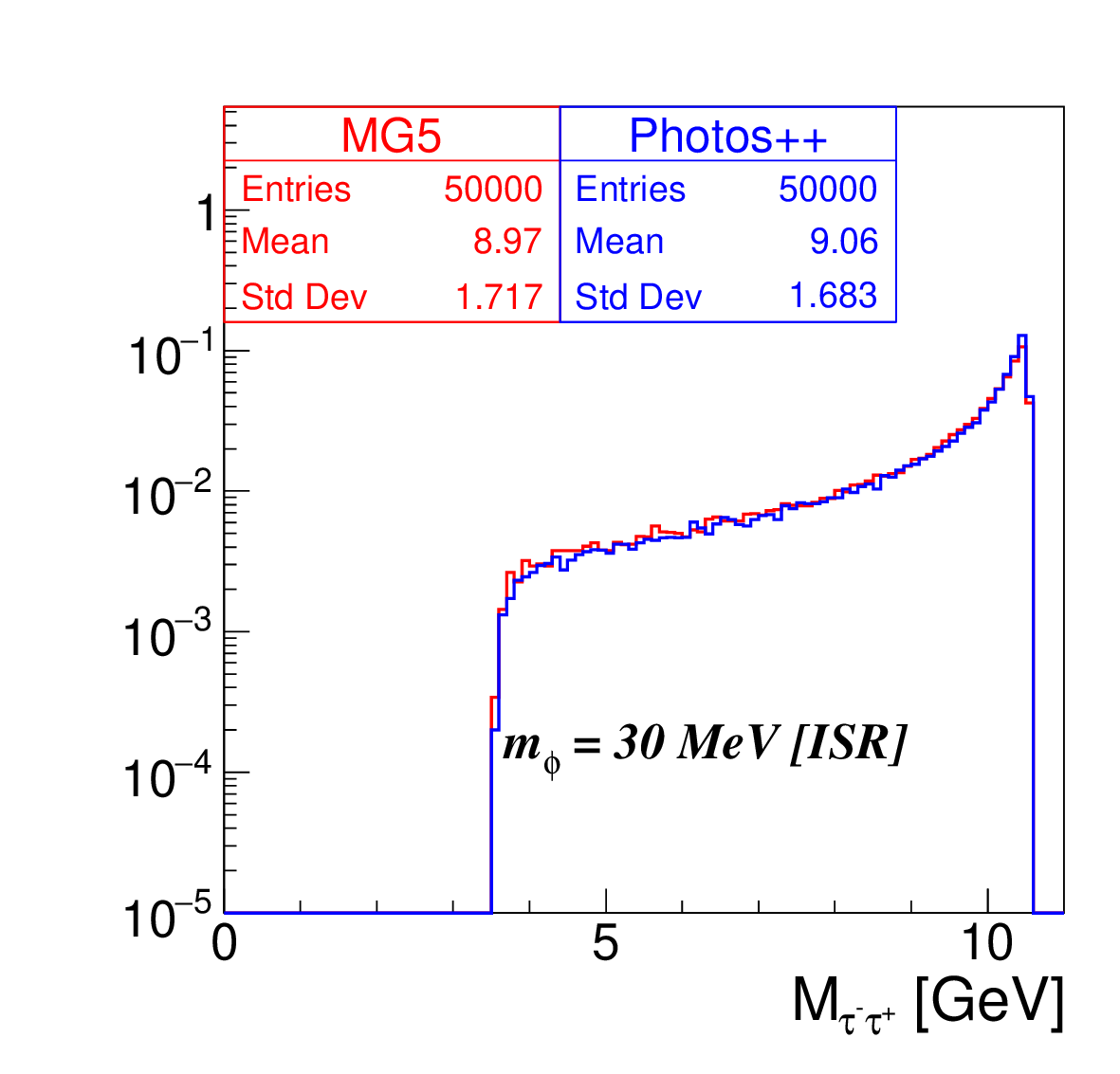}
\includegraphics[width=0.44\textwidth]{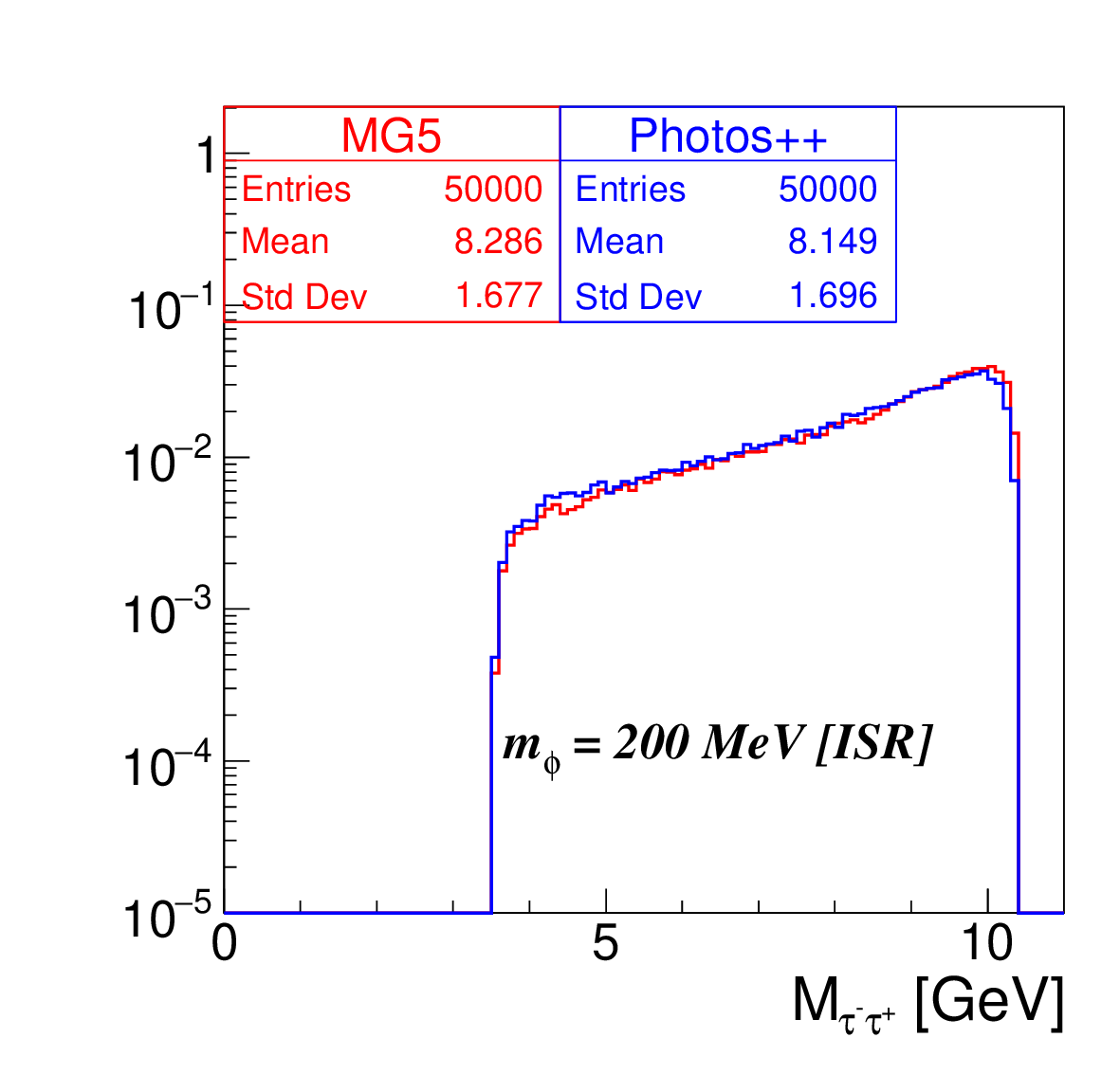}

  \end{center}
  \caption{\small Belle II center of mass energies $e^-e^+ \to \tau^- \tau^+ \phi_{\rm{Dark~Scalar}} (\to e^-e^+)$. Case of dark scalar of 30 and 200 MeV. Simulation of {\tt KKMC+PHOTOS} is compared with the one based on {\tt  MadGraph}.
{Emission kernel was inspired from that comparison.
  At start,  QED pair emission kernel was used. Spin correlations of $\tau$-s
  modified by rotation of $\tau^-$ decay products. Plots taken from \cite{Was:2022nfx}.}}\label{fig:ee}
 \end{figure}

For pair emissions in final states modified algorithm of {\tt PHOTOS} Monte
Carlo can be used. In this case 
phase space is treated exactly, and  no approximations are used, independently
of whether the  pair emission is based on QED calculated matrix element or
of New Physics model. As effects of pair
emissions are  not large, matrix elements are simplified
(in improvable way though)!
Tests/comparisons with matrix element based calculations are essential part of the work. As an example Fig.~\ref{fig:ee} is provided.
    
\section{Anomalous dipole moments}
One of the important aspects of Monte Carlo simulations is the possibility
to imprint into precision event samples generated from Standard Model
predictions, effects of New Physics.
Such effects, which are expected to be small or non-existent were introduced
into event re-weighting algorithms of {\tt TauSpinner} \cite{Przedzinski:2018ett} for  embedding of
anomalous dipole moment  into $pp \to \tau^+\tau^- X$  simulated samples.
Another algorithm for weight imprinting  simultaneously with run of {\tt KKMC}
Monte Carlo for $e^+e^- \to l^+ l^- (n\gamma)$ at low energies and also
for FCC center of mass energies is also  available. See Ref.~\cite{Banerjee:2023qjc}.

For the simulation of New Physics effects, simplified kinematic for  implementation is usually sufficient. There can be two weights calculated,  
 for the cross-section:
 \[ 
wt_{ME}= \Bigl(\sum_{spin }|{\cal M}^{prod\; SM+ NP}|^2 \Bigr) / \Bigl(\sum_{spin }|{\cal M}^{prod\; SM}|^2 \Bigr)
\]

and for the spin effects related to $\tau$ lepton decays. The spin weight
depends on production and on the decay of $\tau$ leptons as well:
 \[ 
 wt_{spin} =\Bigl(\sum_{ij} R^{SM+NP}_{ij} h^i_+ h^j_-\Bigr) / \Bigl(\sum_{ij} R^{SM}_{ij} h^i_+ h^j_-\Bigr)
\]

The $R_{ij}$ depend on the kinematic of $\tau$-pair production, $h^i_\pm$ on $\tau^\pm$ decays.
{
Spin quantization frames orientation needs care. It must be
the same for production and decay.
}
We use {\tt KKMC} routines to transfer $h^i_\pm$ to lab frame and another
routines to transfer back to $\tau^\pm$ rest frame but oriented as in New Physics
calculation.
In this way reference frames are easier to control and impact of photons on
phase space parametrization is not of  great complication.
Note that the solution, if weights are calculated during {\tt KKMC} run, 
is valid  for all $\tau$ decays, while in case of {\tt TauSpinner}, only for main
$\tau$ decay channels spin effects are included. 

In { \tt KKMC} refined solution is used
   for compatibility with  Kleiss-Stirling spinor techniques.
  See Ref.~\cite{Jadach:1998wp}
   for details.
The idea was to relate $\tau$ leptons quantization frames used in production
  and decay through consecutive transformation from $\tau$ rest frame to lab
  frame and back to $\tau$ rest frame.
 That sounds complicated for simple rotation representation, but is safe and
  independent of number of bremsstrahlung photons.
The $\tau^\pm$ decay products and its $h^i_{\tau^\pm}$ vector is
  transformed to the laboratory frame. Then $h^i_{\tau^\pm}$ is transformed back
  to $\tau$ lepton rest-frame, but this time the axes are oriented as chosen
  in Kleiss-Stirling spinor techniques.

  We have explored partly this solution for { \tt KKMC} anomalous moment event re-weighting.
Use of host program frames is convenient but not essential. It helps to improve
 precision, there is no need to worry about bremsstrahlung impact etc. Use of
  internal program variables simplifies the tasks too.
  On the other hand, this  prevents re-use  of events for several distinct models,
  chosen after event samples are already stored in the data files.

\begin{figure}[!htb]
\begin{center}
\includegraphics[width=0.44\textwidth]{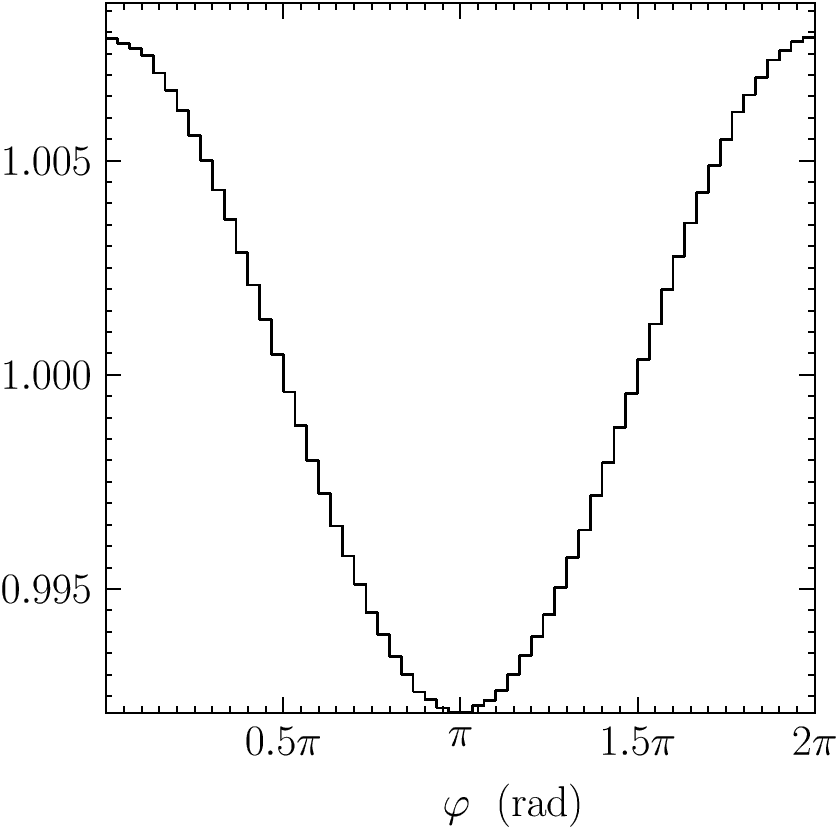}
\includegraphics[width=0.44\textwidth]{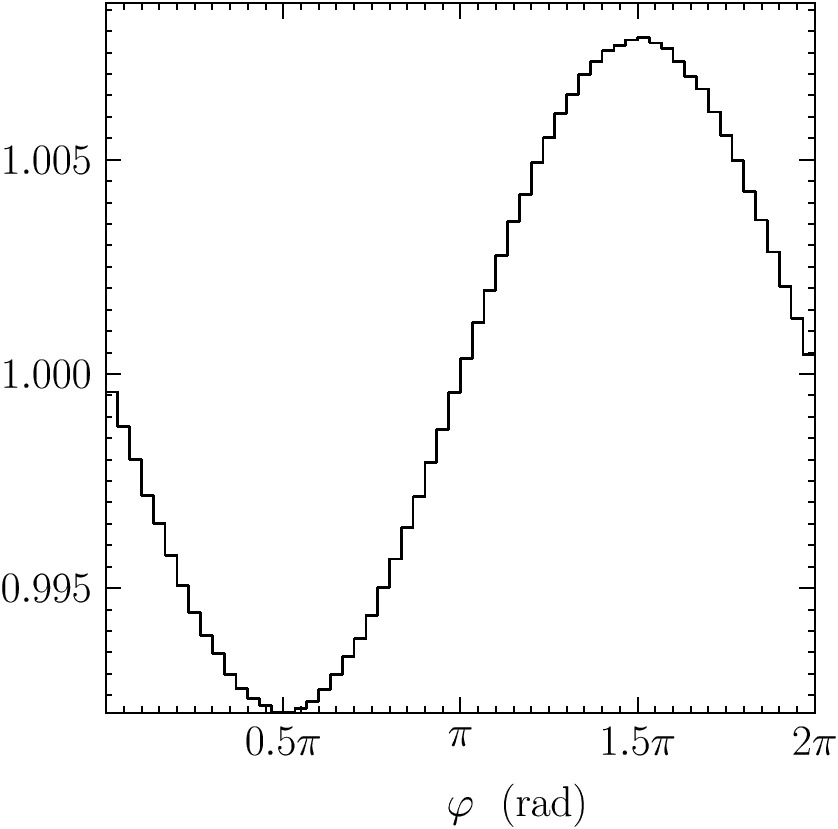} \\
\includegraphics[width=0.44\textwidth]{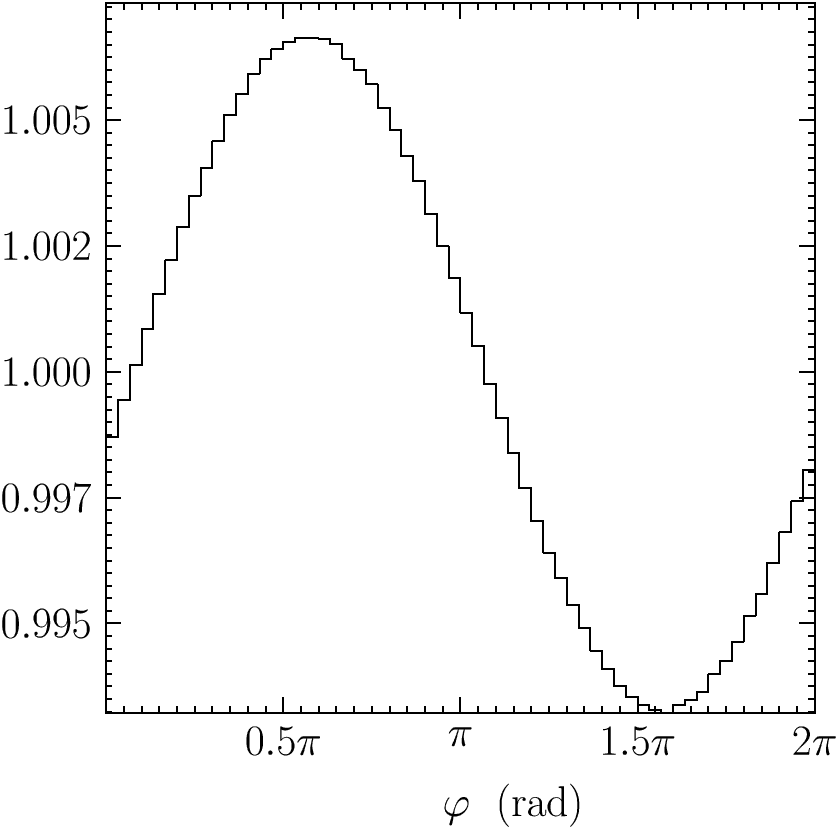}
\includegraphics[width=0.44\textwidth]{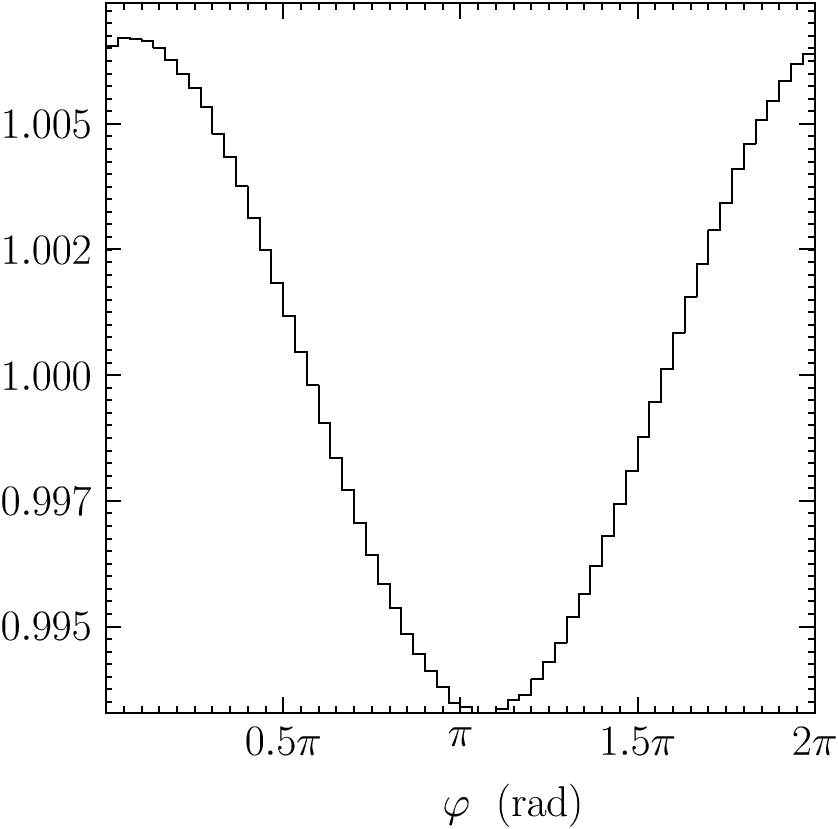}
\end{center}
\centering
\caption{\scriptsize Ratio of number of events with and without weak dipole moments, 
in function of acoplanarity  $\varphi$ at  $\sqrt{s}=M_Z$.  
The selected events of scattering angles  $\cos(\theta)<0$ are taken. 
The top left plot for ${\rm Re}(X) = 0.0004$, the top right plot  for ${\rm Re}(Y) = 0.0004$, 
the bottom left  for  ${\rm Im}(X) = 0.0004$, and the bottom right   for ${\rm Im}(Y) = 0.0004$ are taken . 
For the imaginary form-factors, additional  constraint  $E_{\pi^+} > E_{\bar{\nu}_\tau}$  
is taken  in the $\tau^+$ side.   
The form-factors $A(M_Z^2)=B(M_Z^2)=0$ are set. The decays $\tau^- \to \pi^- \pi^0 \nu$ and $\tau^+\to \pi^+ \nu$ are taken. Plots taken from Ref.~\cite{Banerjee:2023qjc}.}
\label{fig:acoplanarity}
\end{figure}

  The main achievements of the last year include:
  for {\tt KKMC} usage in FCC, extension of re-weighting algorithm to FCC center of
  mass energies, and electroweak corrections are then included, and 
  for  {\tt TauSpinner} usage in LHC, $\gamma \gamma$ parton level processes added, and
  explicit spin correlation matrix $R_{ij}$ prepared for
  quark initialized processes as well. See Ref.~\cite{Banerjee:2023qjc}.
  With the help of {\tt TauSpinner}, semi-realistic observables, see e.g. Fig.~\ref{fig:acoplanarity}, have been
  studied. Later, it is straightforward to move to more realistic ones.
  
\section{Conclusions and outlook }
I have presented some new developments and  and new applications for  Monte Carlo programs family arranged around KKMC. 
Let me finally point to some possible, near future,  activities.
The big issue important for systematic errors are ambiguities 
of $\alpha_{QED}(s)$, the function of electromagnetic coupling. For the Monte Carlos, its  values
for  $s$ starting from $0$ up  to beyond $m_Z^2$ are needed.
Also,  the g-2 measurements revived interest in 
  possible signatures with $\tau$ leptons:  for real productions or  through loops effects. 
  I have presented some work relevant for that. 
  But what about related systematic ambiguities?

 Can modified {\tt KKMC} be helpful in such studies?
  For $\alpha_{QED}(s)$ dispersion relations and measurements of
  $e^+e^- \to \pi^+\pi^-\pi^0 (\pi^+\pi^-)$  at $\sqrt{s} \sim 1$ GeV$^2$ are necessary. This energy range can be
  achieved with Belle II data and events with radiative return. The invariant mass of the hadronic final state
  can be small if it is accompanied by hard initial state bremsstrahlung photon. 
  The question arise if  {\tt KKMC} can be adapted and become helpful.
    
  The necessary steps are as follows.
  Compare program predictions for $e^+e^-\to \mu^+\mu^- \gamma n(\gamma)$ at   {\tt (C)EEX1} and {\tt (C)EEX2}
  levels.
  Replace $\mu^\pm$ with $\pi^\pm$ in {\tt KKMC}  generation with QED initial state bremsstrahlung only.
  Use ratio of $e^+e^-\to \pi^+\pi^- \pi^0$ to $e^+e^-\to \mu^+\mu^-$  matrix elements at interpolated phase space points
  of $2 \to 2$ kinematic, to calculate event weight.
  Imprint into final state $\pi^0$ using modified {\tt PHOTOS} Monte Carlo, and make tests with alternative
  simulations of    $e^+e^-\to \pi^+\pi^- \pi^0$ process (no bremsstrahlung photons).
   Reproduce old tests with {\tt PHOKARA} \cite{Rodrigo:2002hk,Czyz:2014sag}, see. e.g.
   {\tt https://indico.ph.liv.ac.uk/event/1297/contributions/7323/}~. Unfortunately authors of that works
   are not any more available to help.

  I have not presented  further details on 
   arrangements for $\tau$ decays and 
  for software, except what was explained in the introduction.

  \section*{Acknowledgments}
  \paragraph{Funding information}
This research was funded in part by ``National Science Centre, Poland'', grant no.  2023/50/A/ST2/00224.




\bibliography{ZWasLoui.bib}

\nolinenumbers

\end{document}